\documentclass[acmlarge,nonacm]{acmart}
\AtBeginDocument{%
  \providecommand\BibTeX{{%
    \normalfont B\kern-0.5em{\scshape i\kern-0.25em b}\kern-0.8em\TeX}}}

\setcopyright{acmcopyright}
\copyrightyear{2018}
\acmYear{2018}
\acmDOI{10.1145/1122445.1122456}

\acmJournal{POMACS}
\acmVolume{37}
\acmNumber{4}
\acmArticle{111}
\acmMonth{8}

\usepackage{graphicx}
\usepackage{caption}              
\usepackage{subfig}
\usepackage{float}

\begin{document}

\title{Capitol (Pat)riots: A comparative study of Twitter and Parler}


\author{Hitkul}
\affiliation{%
  \institution{IIIT - Delhi}
  \country{India}}
\email{hitkuli@iiitd.ac.in}

\author{Avinash Prabhu}
\authornote{Authors contributed equally to this research.}
\author{Dipanwita Guhathakurta}
\authornotemark[1]
\author{Jivitesh jain}
\authornotemark[1]
\author{Mallika Subramanian}
\authornotemark[1]
\author{Manvith Reddy}
\authornotemark[1]
\author{Shradha Sehgal}
\authornotemark[1]
\author{Tanvi Karandikar}
\authornotemark[1]

\affiliation{%
  \institution{IIIT - Hyderabad}
  \country{India}
}

\author{Amogh Gulati}
\authornotemark[1]
\author{Udit Arora}
\authornotemark[1]
\affiliation{%
  \institution{IIIT - Delhi}
  \country{India}
}

% \author{Jushaan Singh Kalra}
% \affiliation{%
%   \institution{Delhi Technological University}
%   \country{India}
% }

\author{Rajiv Ratn Shah}
\affiliation{%
  \institution{IIIT - Delhi}
  \country{India}}
\email{rajivratn@iiitd.ac.in}

\author{Ponnurangam Kumaraguru}
\affiliation{%
  \institution{IIIT - Delhi}
  \country{India}}
\email{pk@iiitd.ac.in}

\renewcommand{\shortauthors}{Hitkul, et al.}

\begin{abstract}
  On 6 January 2021, a mob of right-wing conservatives stormed the USA Capitol Hill interrupting the session of congress certifying 2020 Presidential election results. Immediately after the start of the event, posts related to the riots started to trend on social media. A social media platform which stood out was a free speech endorsing social media platform Parler; it is being claimed as the platform on which the riots were planned and talked about. Our report presents a contrast between the trending content on Parler and Twitter around the time of riots. We collected data from both platforms based on the trending hashtags and draw comparisons based on \textit{what} are the topics being talked about, \textit{who} are the people active on the platforms and how \textit{organic} is the content generated on the two platforms. While the content trending on Twitter had strong resentments towards the event and called for action against rioters and inciters, Parler content had a strong conservative narrative echoing the ideas of voter fraud similar to the attacking mob. We also find a disproportionately high manipulation of traffic on Parler when compared to Twitter.  
\end{abstract}



\keywords{Social Computing, Data Mining, Social Media Analysis, Capitol Riots, Parler, Twitter}

\maketitle

\section{Introduction}

Misinformation of the United States of America's presidential election results being fraudulent has been spreading across the world since the elections in November 2020.\footnote{\url{https://www.bbc.com/news/election-us-2020-55016029}} Public protests and legal cases were taking place across the states against the allegation of voter fraud and manhandling of mail-in ballots.\footnote{\url{https://www.business-standard.com/article/international/trump-approaches-us-supreme-court-against-presidential-election-results-120121000240_1.html}} This movement took a violent height when a mob attacked the Capitol Hill building to stop certification of Mr. Joe Biden as the 46th President of the United States of America. The incident lead to a mid-way halt of a running congress session and the death of five people including a police officer.\footnote{\url{https://www.theguardian.com/us-news/2021/jan/08/capitol-attack-police-officer-five-deaths}} The incident also spread its ripples in the online world with hashtags like \#capitolriots that started trending on Twitter, Google search and other social media platforms.  

A multitude of research has been done in uncovering the role of social media in politics from various aspects like misinformation, topical focus of content, communities, and bot accounts \cite{10.1145/3392860,ferrara2017disinformation,morstatter2018alt,bessi2016social}. The USA saw an example of large scale use of social media for political protest earlier in 2020, after the killing of George Floyd incited widespread support for the BlackLivesMatter (BLM) movement on Twitter ~\cite{twitter:blm}. The US 2020 presidential elections also saw a resurgence of the platform with \#Election2020 and \#NovemberIsComing trending, in order to spark election-related conversations ~\cite{twitter:elec}.

% \citet{pol1} and \citet{pol2} points out that Social media plays a crucial role in political events as it makes discourse around politics more accessible and widespread. Social media is often used to mobilize political movements, raise awareness, protest policies, campaign for elections, and to debate political events around the globe. 

In this report, we collect data from two Online Social Media sites (OSMs) - Twitter\footnote{\url{https://www.twitter.com/}} and Parler\footnote{\url{https://parler.com/}} - that were actively used to discuss the riots. Parler is a free speech social network that garnered attention when President Donald Trump publicly denounced social media giants like Twitter and Facebook for targeting him and other conservatives. The network has also been used by right-wing extremists to plan the Jan 6th breach of the Capitol.\footnote{\url{https://www.businessinsider.in/tech/news/plans-to-storm-the-capitol-were-circulating-on-social-media-sites-including-facebook-twitter-and-parler-for-days-before-the-siege/articleshow/80155657.cms}} Section~\ref{parler} goes into details of Parler rules and functioning. 

We conduct a comparative study of the trending content and users on the two platforms. We observe a violent rhetoric on Parler with a significant portion of the content in support of the Capitol riots and misinformed claims of fraudulent elections.\footnote{\url{https://www.thehindu.com/news/international/us-supreme-court-rejects-republican-attack-on-biden-victory/article33312520.ece}} Content on Twitter, by contrast, denounced the storming of the Capitol and the weak response from police to the incident, in contrast to the BLM protests. Users on Twitter shared their concerns regarding the rising violent riots in America. A longitudinal analysis of the content and the users' joining dates highlights how the two OSMs were actively used during and after the protests, reaffirming the point that social media has a significant role to play in political events.

\section{An Introduction to Parler}
\label{parler}
Parler is a micro-blogging and social networking service launched in 2018 with headquarters in Henderson, Nevada, USA. According to the About page, the network is a \textit{free speech social network}, built on a \textit{foundation of respect for privacy and personal data, free speech, free markets and ethical, transparent corporate policy}.\footnote{\url{https://company.parler.com}} Parler advertises the minimal rules and content guidelines it imposes, explaining its popularity amongst users who are banned from popular social media websites such as Twitter and Facebook due to their content moderation policies.\footnote{\url{https://www.nytimes.com/2020/11/11/technology/parler-rumble-newsmax.html}} In the aftermaths of the Capitol attack, Parler got suspended by Amazon Web Services hosting leading to platform being unavailable.\footnote{\url{https://www.cnbc.com/2021/01/11/parler-drops-offline-after-amazon-withdraws-support.html}}

\subsection{Parler Rules}
The Parler Community Guidelines are based on two principles: 1) Parler's services cannot be used as a tool for crime and unlawful acts and 2) Bots and spam are a nuisance and not conducive to productive and polite discourse. While threats of violence and advocacy of lawless actions are prohibited, fighting words and NSFW (Not-Safe-For-Work) content is allowed, under some restrictions.\footnote{\url{https://legal.parler.com/documents/Elaboration-on-Guidelines.pdf}}
Reported violations of these guidelines are reviewed by a Community Jury which determines whether the content is permitted or not. A point system is in place to ban repeated and frequent offenders.\footnote{\url{https://legal.parler.com/documents/Parler-Community-Jury.pdf}}

\subsection{Parler Social Network Structure}

Parler allows registered users to write \textit{parleys}, which are posts at most 1,000 characters long. Social Network engagement features such as  comments and votes on parleys written by others are also present. Each user has their own \textit{feed} - a stream of parleys that they can interact with. Unlike other popular platforms the feed is not curated by Parler. Users curate and moderate their feed using options provided by the platform - reflecting Parler's principles. Parler allows users to search for hashtags and usernames. The lack of search in text leads to an overemphasis on the use of hashtags.

% To set up an account on parler, a user has to choose a name and a username. After setting up an account, a user sends and receives parleys; which are 1000 character posts. Once a parley is posted, it appears in a user's feed as well on the feeds of those who follow the user.

% Searching on parler enables a user to search for terms (min. 3 characters) only in hashtags and username, it is not possible to search for terms in posts. This would severely affect the way users post as including the right hashtags and mentions would be more important that the post content itself.

Another means of exploring content on Parler is through the \textit{discover} section. The discover section consists of parleys, people and \textit{affiliates}. Affiliates are news outlets which are allowed by Parler to post their news articles. Table~\ref{tab:terms} provides a description of common terms and actions associated with the network. 

% \ref{parler:terms}.

\begin{table}
\renewcommand{\arraystretch}{1.2}
\caption{Terminology used by Parler.}
\label{tab:terms}
\centering
\begin{tabular}{ | p{4cm} | p{10cm}|  }
\hline
\textbf{Term} & \textbf{Definition}
\\ 
\hline
Parley & A Parley is a 1,000 word post that can be shared on the Parler Platform.  \\ 
\hline
Hashtag & A word or phrase preceded by a hash sign (\#), used on Parler to identify digital content on a specific topic. \\ 
\hline
Comment & A Comment is a 1,000 character reply to a Parley. \\
\hline
Echo & An Echo is a re-posting of a Parley. Parler’s Echo feature helps users quickly share that Parley with all of their followers. \\
\hline
Vote & A Vote on a Parley is a way to let people know that the user enjoys it. \\
\hline
Direct Message or DM & Users can directly contact anyone by going to their Parler profile and selecting the message icon to start a conversation. \\
\hline
Follow & Following another user means that all their parleys will appear
in the feed.
\\
\hline
Unfollow & Unfollowing another user means that all their parleys will no longer appear
in the feed.
\\
\hline
Citizen & Parler Citizens are verified unique people in the Parler network.
\\
\hline
Verified & People with a large following have the potential to be targeted for impersonation, hacking or phishing campaigns. The verified badge is given to protect the person’s authenticity and prove their identity to the community.
\\
\hline
Block & If one user blocks another, he/she won't be be able to see the blocked account and vice versa.
\\
\hline
Mute & Muting a user will prevent the users’ posts from showing up on the feed.
\\
\hline
\end{tabular}
\end{table}

% \subsection{Recent Rise in Popularity and Connection to the Event}
% Founded in 2018, Parler started off as a small player in the social media landscape but has seen a huge surge in activity in the past year. The site has been embraced by prominent voices on the right, from Fox News commentator Sean Hannity to U.S. Sen. Ted Cruz, R-Texas. Users herald the self-described free speech social network for its hands-off approach to moderating content. The site gained lot of traction during the \textit{#twexit} movement started by the founders of Parler when President Donald Trump went to war with Social Media Giants (May-July 2020). The site also gained a lot of users in and around the election time (Nov 2020). The site currently hosts 4 million active users and 10 million users in total. 

% Parler is claimed to be a radical Right-wing social media website and one of the tools used in planning the storming of the Capitol Building on January 6th 2021 and we wish to study this claim in this Report.

\section{Data Collection}
Our data collection was done on the 7 and 8 January 2021. A list of trending hashtags and keywords were curated, and used as the seed to collect data. Parleys were collected using parler-py-api. \footnote{\url{https://github.com/KonradIT/parler-py-api/}} Data collected was dated between 1 Nov 2020 03:39 am EST to 8 Jan 2021 08:15 am EST. A total of approximately 100,000 parleys from 22,000 unique users were collected.\footnote{\url{http://precog.iiitd.edu.in/resources.html}}

For collecting Tweets, we used the official Twitter Streaming API. Our collection period was  7 Jan 8:00 AM EST to 8 Jan 7:15 AM EST, while adhering to rate limits. A total of approximate 4 Million tweets were collected from 1.7 Million users. Table~\ref{tab:data_statsa} provides a summary of our dataset statistics. 

\begin{table}[]
\caption{Summary of dataset statistics}
\label{tab:data_statsa}
\begin{tabular}{|l|r||l|r|}
\hline
\multicolumn{2}{|c||}{\textbf{Parler}} & \multicolumn{2}{c|}{\textbf{Twitter}} \\ \hline
Total Parleys        & 101,945        & Total Tweets        & 4,196,988       \\ \hline
Echos                & 35,165         & Retweets            & 3,288,274       \\ \hline
Unique users         & 22,326         & Unique users        & 1,720,826       \\ \hline
\end{tabular}
\end{table}

\section{A Comparative Study}
We compared the collected data from Parler and Twitter on the basis 1)\emph{What} was being posted, 2)\emph{Who} were the people posting and creating engagement, and lastly 3)How \emph{organic} was the traffic generated. 

\subsection{What was being posted?}
\begin{figure}[!tbp]
  \centering 
  \def\svgscale{0.5}
  \subfloat[Parler]{\includegraphics[width=0.475\textwidth]{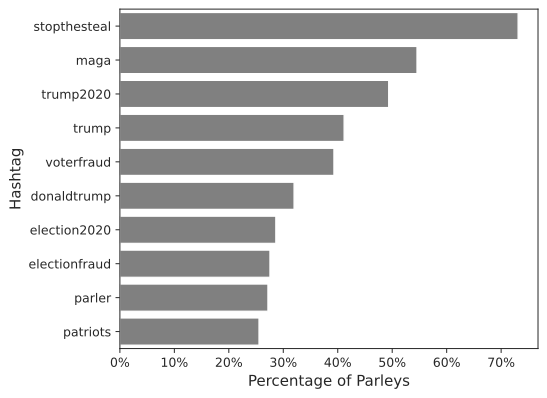}\label{fig:p_hash}}
  \hfill
  \subfloat[Twitter]{\includegraphics[width=0.525\textwidth]{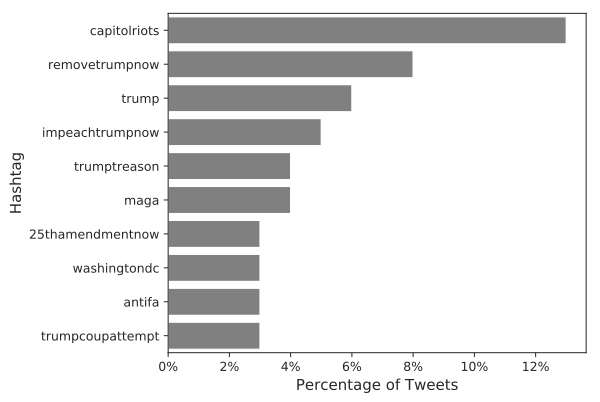}\label{fig:t_hash}}
  \caption{Ten Most frequently used hashtags. Hashtags on Parler are echoing the ideas similar to the Capitol attacking mob, Twitter is treding with hashtags representing a call for action against the event.}
  \label{fig:hash}
\end{figure}

To understand each platform's topics of trending conversation, we looked at the top ten most common hashtags on respective platforms. Figure~\ref{fig:hash} shows the ten most frequently used hashtags represented with the percentage of post they appear.  Percentage of use for a particular hashtag has been calculated by counting the number of posts mentioning that hashtag at least once, normalised by the total number of posts on that platform containing at least one hashtag. 

We observe a stark contrast in the hashtag popularity on two platforms. All the hashtags trending on Parler represent the misinformed idea of voter fraud and echo the ideas similar to attacking mob. Hashtags popular on Twitter are either neutral towards the event, e.g. \emph{\#capitolriots} and \emph{\#washingtondc} or the ones who are calling out for impeachment of President Donald Trump who is being called responsible for inciting his follower for this attack.\footnote{\url{https://www.washingtonpost.com/politics/trump-rage-riot/2021/01/07/26894c54-5108-11eb-b96e-0e54447b23a1_story.html}} One exception to this result is the presence of \emph{\#maga} on Twitter, which is associated with the President Trump's election campaign. 

It is also worth noting that most frequent hashtag on Parler is present in above 70\% of posts compared to only above 12\% in case of Twitter representing an extreme one-sided narrative present on Parler. To provide a better insight of context in which these hashtags have been used, Table~\ref{tab:p_tag_context} and Table~\ref{tab:t_tag_context} shows 5 parleys and tweets for two most frequent hashtags respectively.

\begin{table}[]
\caption{Sample Parleys using popular hashtags.}
\label{tab:p_tag_context}
\renewcommand{\arraystretch}{1.2}
\begin{tabular}{|p{7.5cm}|p{7.5cm}|}
\hline
\multicolumn{1}{|c|}{\textbf{\#stopthesteal}}                                                                                                                          & \multicolumn{1}{c|}{\textbf{\#maga}}                                                                                                                                                                                                                                                                                                         \\ \hline
Interesting.. The cops should have just went home.. We will never call Sleepy Creepy PEDO  maker Joe our President. \#DrainTheSwamp \textbf{\#StopTheSteal} \#FAKENEWSROBOTS \#StillYourPresident \#FREEDOM & January 21st, 2021 we rise up again...I CALL BULLSHIT \#JOEBIDEN is not  my president!...\textbf{\#maga}                                                                                                                                                                                                                                                \\ \hline
...they stopped counting in Chatham County at like 10:30-11 last night - "resumed" this morning. Republicans  were leading both seats with over 90\% of the vote in...\textbf{\#stopthesteal}...                                                                                                                  & MAGA PATRIOTS please support Trump and storm the Capitol and Congress properly.!! We will install Truml as our Lord and savior, PRESIDENT FOR LIFE... because we won the election!! THIS TIME WE TAKE THE CAPITOL WITH GUNS bcs it is not fair we lost DESTORY AMERICA OR TRUMP STAYS PRESIDENT... its the most patriotic thing to do...\textbf{\#maga} \\ \hline
Antifa is in there. They have breached the House...\textbf{\#StopTheSteal} \#voterfraud...          & DO THE MATH!...\textbf{\#maga}                                                                                                                                                                                                                                                                                                                          \\ \hline
75\% \#CCP controlled voting machines used in United States election destroys trust in \#electionintegrity ! \textbf{\#StopTheSteal}...                                                                     & ...Treason against the United States is taking place surrounding the presidential election. The largest cyber warfare activity in the world...\textbf{\#maga}...                                                                                                                                                                                        \\ \hline
Still haven't seen where Biden and Harris have taken their vaccines. Has anyone else seen it?...\textbf{\#StopTheSteal}                                                                                     & Despite the crisis, a sense of unity \& patriotism afford \#America a common mission \& increased opportunities...\textbf{\#maga}...                                                                                                                                                                                                                    \\ \hline
\end{tabular}
\end{table}

\begin{table}[]
\caption{Sample Tweets using popular hashtags.}
\label{tab:t_tag_context}
\begin{tabular}{|p{7.5cm}|p{7.5cm}|}
\hline
\multicolumn{1}{|c|}{\textbf{\#capitolriots}}                                                                                                                          & \multicolumn{1}{c|}{\textbf{\#removetrumpnow}}                                                                                                                                                                                                                 \\ \hline
Dear @VP @Mike\_Pence: In light of the \textbf{\#CapitolRiots} yesterday, there are bipartisan calls for you to invoke the 25t… & President and Commander in Chief Trump should be either removed from office with the 25th amendment, impeached, and/or investigated for criminal charges. \#ImpeachTrumpNow \textbf{\#RemoveTrumpNow} \#InvestigateTrump \\ \hline
Confirmed that some police were involved in the Capitol attack yesterday \textbf{\#CapitolRiots}                                & \textbf{\#RemoveTrumpNow} He betrayed the country he swore to protect. The Constitution and Democracy mean nothing to him. He…                                                                                           \\ \hline
The \textbf{\#CapitolRiots} yesterday were underpinned by pure racist hatred nothing less. This started with Trumps birtharis…  & We can’t wait 13 days. \textbf{\#RemoveTrumpNow}                                                                                                                                                                         \\ \hline
\#BREAKING: In response to Trump inciting the deadly                                                                                             & People have died because of Trump's incitement and sedition. \textbf{\#RemoveTrumpNow}                                                                                                                                   \\ \hline
The \textbf{\#CapitolRiots} were a terrorist act incited by Donald Trump, Don Jr, Rudy Giuliani and members of Congress. Arre…  & The right time to do the right thing is RIGHT NOW.  \textbf{\#RemoveTrumpNow}                                                                                                                                            \\ \hline
\end{tabular}
\end{table}
%%%%%%%%%%%%%%%%%%%%%%%%%%%%%%%%%%%%%%%%%%%%%

\begin{figure}[!tbp]
  \centering 
  \def\svgscale{0.5}
  \subfloat[Parler]{\includegraphics[width=0.485\textwidth]{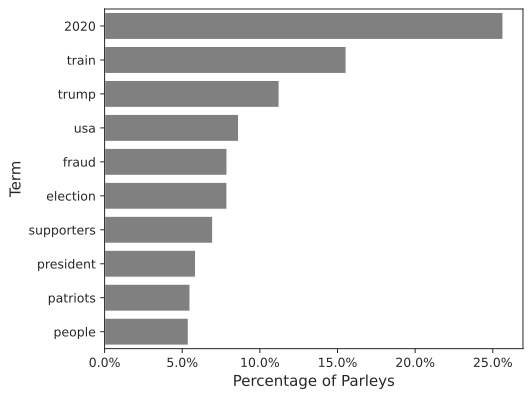}\label{fig:p_term}}
  \hfill
  \subfloat[Twitter]{\includegraphics[width=0.5\textwidth]{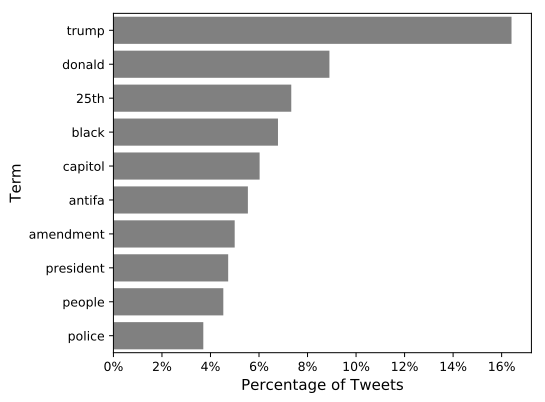}\label{fig:t_term}}
  \caption{Ten most frequently used terms. Conservative and election fraud terms most frequently appeared on Parler representing an association with the idea of rioters. However, Twitter is dominated by call for 25th amendment and comparisons with black live matter protest.}
  \label{fig:term}
\end{figure}

To get further insights into the content being posted on both platforms, we repeated the frequency plots for terms shown in Figure~\ref{fig:term}. We have used the word \textit{term} in this context to mean any uni-gram (excluding stop words) included in a post, but not as a mention of a user or a hashtag. Percentage of use for a particular term has been calculated by counting the number of posts mentioning that term at least once, normalised by the total number of posts on that platform. We observe the same pattern as shown by hashtags, terms used on Twitter, indicating a sense of dissatisfaction and disdain towards the US Capitol's actions. On the other hand terms, frequent on parley indicate a strong sense of support towards undermining the veracity of the 2020 US presidential elections. Parleys also display abundant use of strong language compared to the tweets, which can be easily attributed to the liberal community guidelines, and supports claims that the platform is used as a medium for proliferating violent and offensive content.

Apart from this stark difference in opinion, some terms harbour a deeper meaning. On Twitter, terms like \textit{25th Amendment} advocate President Donald Trump's immediate removal from office. In contrast, those like \textit{black} draw a comparison between law enforcement agencies' response to the demonstrations at the US Capitol and the Black Lives Matter protests of July 2020, adding a racial dimension to the conversation.

Since terms \textit{trump} and \textit{president} are common and occur frequently in both the platforms, we compare the contexts in which two popular terms appear in parleys and tweets in Table~\ref{tab:trump_context} and Table~\ref{tab:president_context}. These Tables reinforce our earlier inferences of Parler content supporting the riot and voter fraud narrative, whereas Twitter samples are against the riot and President Trump.

\begin{table}[]
\caption{Context Comparison for the term \textit{Trump}.}
\label{tab:trump_context}
\begin{tabular}{|p{7.5cm}|p{7.5cm}|}
\hline
\multicolumn{1}{|c|}{\textbf{Parler}}                                                                                                                                                                                                                                                                                                                                                   & \multicolumn{1}{c|}{\textbf{Twitter}}                                                                                                                                          \\ \hline
Plandemic Exposed...\textbf{Trump} knows that the Dummies are already forging their narrative with how they will stop this plandemic as soon as Chinese Joe gets in the Oval Office...                                                                                                                                                           & Well, \textbf{Trump} finally got his wall. It’s in Washington DC.                                                                        \\ \hline
...I urge all \textbf{Trump} supporters and lovers of freedom to protest at your place of government. We can not let this nation fall to the hands of foreign puppets...                                                                                                                                                                         & No matter how \textbf{Trump} now behaves...                                                                                              \\ \hline
Listening to \textbf{Trump} speak. \#stopthesteal                                                                                                                                                                                                                                                                                                & Donald \textbf{Trump} needs to resign or be removed from office. America has endured enough.                                             \\ \hline
\textbf{Trump} pledging appear personally at the rally of outright fascists and white supremacists on January 6th as part of his coup efforts. This is an act of intimidation that has parallels with Mussolini's march on Rome. It is aimed at intimidating not just the working class as a whole but Trump's opponents...\#fascsim \#Socialism & \textbf{Trump} did NOT immediately send out the National Guard. No initial presence. Clashes began $\sim$1:20pm. The breach happ...      \\ \hline
I'm not saying there was no vandalism or violence from our side but I didn't see any of it for the 8+ hours I was there. I did watch ANTIFA breaking windows at the Capitol and a group of \textbf{Trump} supporters tackled him in 30 seconds. \#StopTheSteal                                                                                   & ...Speaker Nancy Pelosi said Congress will IMPEACH \textbf{trump} if he's not immediately removed by the 25th Amendment. LET'S GOOOO!!!! \\ \hline
\end{tabular}
\end{table}

\begin{table}[]
\caption{Context Comparison for the term \textit{President}.}
\label{tab:president_context}
\begin{tabular}{|p{7.5cm}|p{7.5cm}|}
\hline
\multicolumn{1}{|c|}{\textbf{Parler}}                                                                                                                                                                                                                                                                              &\multicolumn{1}{c|}{\textbf{Twitter}}                                                                                                                                  \\ \hline
...It took a small group of Patriots to enter the Capitol Building to get the attention of America about the \#voterfraud that occurred that made Joe Biden a fraudulent \textbf{president} elect. Almost 50\% of Americans do not think it was a free and fair election... & Why are these swamp creatures in such a rush to use the 25th amendment or to impeach \textbf{President} @realDonaldTrump? Do…   \\ \hline
I proudly stand with \textbf{President} Trump and I will fight to the death to ensure the sanctity of our Republic...                                                                                                                                                       & The Vice \textbf{President} and the Cabinet should vote, today, on invoking the 25th amendment. Every second that Donald Tr...  \\ \hline
\textbf{President} \#DonaldTrump signed an executive order on Nov. 12 prohibiting Americans from investing in select Chinese firms that support China.                                                                                                                      & All that these cabinet resignations say to me is that they don’t have the guts to hold \textbf{President} Trump accountable...  \\ \hline
GOD BLESS \textbf{PRESIDENT} TRUMP!!! The \textbf{president} says HELL NO! to the PORK STUFFED BURRITO that CONGRESS calls A STIMULUS BILL!!!...\#BIDENCRIMEFAMILY..                                                                                       & NEW VIDEO: The \textbf{President} has betrayed the country and incited an insurrection against our own government. Retweet i... \\ \hline
I also stand with \textbf{President Trump}...\#donaldtrumpismypresident \#trump4moreyears...                                                                                                                                                                                & Donald Trump will be remembered as the worst \textbf{president} in the history of the United States.                            \\ \hline
\end{tabular}
\end{table}

\subsection{Who were the people posting?}
Users on both platforms were compared across three attributes - Content generated, Mentions and Reposts. Since the datasets are vastly different in size, we normalize by number of posts to gauge the differences between the platforms. We start with filtering out users by the volume of content they generate. We also compare user bios and joining date to get a bird's eye view on the users' type present on the platforms. Tables~\ref{parler:active} and \ref{twitter:posts} shows the five most active users on Parley and Twitter respectively.

\begin{table}
\parbox{.45\linewidth}{
\centering
\renewcommand{\arraystretch}{1.2}
\caption{Most Active Users on Parler. Five users account for 11\% content generated.}
\label{parler:active}
\centering
\begin{tabular}{ | p{3.2cm} | p{1.5cm}| p{2cm} | }
\hline
\textbf{User} & \textbf{Posts} & \textbf{Percentage}
\\ 
\hline
Patriots4US & 6320 & 6.19\%  \\ 
\hline
TheRealWakeUpMfers & 2580 & 2.53\%  \\
\hline
GameOver & 1406 & 1.37\%  \\
\hline
Billyboy428 & 1105 & 1.08\%  \\
\hline
marylandcrabbing & 994 & 0.97\%  \\
\hline
\end{tabular}
}
\hfill
\parbox{.45\linewidth}{
\centering
\renewcommand{\arraystretch}{1.2}
\caption{Most Active Users on Twitter.}
\label{twitter:posts}
\begin{tabular}{ | p{2.5cm} | p{1.5cm}| p{2cm} | }
\hline
\textbf{User} & \textbf{Posts} & \textbf{Percentage}
\\ 
\hline
openletterbot & 511 & 0.0001\%  \\ 
\hline
4Tchat & 360 & 0.0001\%  \\
\hline
Difference30360 & 333 & 0.0001\%  \\
\hline
Anime\_ABEMA & 330 & 0.0001\%  \\
\hline
RogueRiverSun & 317 & 0.0001\%  \\
\hline
\end{tabular}
}
\end{table}

We once again observe a large contrast between the users and their posts. The users on Parler posted many extreme Parleys in which they urged people to take part in the protest and later showed support for the protest. On the other hand, Twitter users \textit{openletterbot} and \textit{RogueRiverSun} posted extensively against the protest. \textit{openletterbot} is a bot used to deliver messages from the public to elected officials and posted many anti-protest messages during the Capitol storming. On Parler approximately 11\% of the total content is generated by only five users this may indicate that there is heavy traffic manipulation on Parler, we attempt to study this in Section~\ref{ctm}.

Next, in Table~\ref{parler:mentions} and Table~\ref{twitter:mentions} we list the top five most mentioned users on both platforms. We observe President Trump's account getting most mentions on both the platforms. However, there is a stark contrast for the rest of the users on the list. We observe predominantly right-wing personalities being mentioned frequently on Parler whereas on Twitter, we observe a larger inclination to left-wing leaders being mentioned. The lack of left-wing leaders not being mentioned on Parler may be due to the fact that most left-Wing Leaders do not have a Parler account. The FBI is also mentioned dominantly on Twitter due to a movement in which the identity of the people involved in the riot was identified on Twitter and reported to the FBI account.\footnote{\url{https://twitter.com/FBI/status/1348283582490546177?s=20}}

\begin{table}
\parbox{.45\linewidth}{
\centering
\renewcommand{\arraystretch}{1.2}
\caption{Most Mentioned Users on Parler. Predominately right-wing user accounts. }
\label{parler:mentions}
\centering
\begin{tabular}{ | p{2.5cm} | p{1.5cm}| p{2cm} | }
\hline
\textbf{User} & \textbf{Mentions} & \textbf{Percentage}
\\ 
\hline
TeamTrump & 3363 & 3.30\%  \\ 
\hline
linwood & 2802 & 2.75\%  \\
\hline
SeanHannity & 2185 & 2.14\%  \\
\hline
Marklevinshow & 2139 & 2.09\%  \\
\hline
GenFlynn & 1805 & 1.77\%  \\
\hline
\end{tabular}
}
\hfill
\parbox{.45\linewidth}{
\centering
\renewcommand{\arraystretch}{1.2}
\caption{Most Mentioned Users on Twitter. Predominately left-wing leader, President Trump being an exception. The movement of Identify rioter. online led to heavy presence of FBI. }
\label{twitter:mentions}
\begin{tabular}{ | p{2.5cm} | p{1.5cm}| p{2cm} | }
\hline
\textbf{User} & \textbf{Mentions} & \textbf{Percentage}
\\ 
\hline
realDonaldTrump & 87223 & 0.023\%  \\ 
\hline
AOC & 53418 & 0.014\%  \\
\hline
FBI & 37317 & 0.010\%  \\
\hline
HawleyMO & 32931 & 0.009\%  \\
\hline
SpeakerPelosi & 30325 & 0.008\%  \\
\hline
\end{tabular}
}
\end{table}

In Tables~\ref{repost} and \ref{twitterrepost} we list top five users with the highest number of Reposts on both platforms. Most of the accounts belonged to people who were involved in Media and Press. However, the content that two sets of users posted were extensively different. A majority on Parler posted in support of the ``revolution" and believed strongly in the riots whereas the users on Twitter put up posts which condemned the riot on the platform. This strongly shows the disparity in the content that is spreading on both websites. We further notice an abnormally high repost percentage on Parler coming from a small set of users, this combined with the single-sided narrative of most user-generated content is a vital sign of a platform level echo chamber on Parler.  

\begin{table}
\parbox{.45\linewidth}{
\centering
\renewcommand{\arraystretch}{1.2}
\caption{Most Reposted Users on Parler. Twenty percent of repost content being generated by five conservative accounts, present strong sign of skewed narrative on Parler.}
\label{repost}
\centering
\begin{tabular}{ | p{2.7cm} | p{1.5cm}| p{2cm} | }
\hline
\textbf{User} & \textbf{Reposts} & \textbf{Percentage}
\\ 
\hline
WarRoomPandemic & 11352 & 11.1\%  \\ 
\hline
Ryanahlberg & 5342 & 5.2
\%  \\
\hline
epochtimes & 2017 & 1.9
\%  \\
\hline
tjf2020 & 1538 & 1.5
\%  \\
\hline
JoePags & 1123 & 1.1\%  \\
\hline
\end{tabular}
}
\hfill
\parbox{.45\linewidth}{
\centering
\renewcommand{\arraystretch}{1.2}
\caption{Most Retweeted Users on Twitter. Predominantly left-wing accounts were retweeted on Twitter.}
\label{twitterrepost}
\begin{tabular}{ | p{2.5cm} | p{1.5cm}| p{2cm} | }
\hline
\textbf{User} & \textbf{Reposts} & \textbf{Percentage}
\\ 
\hline
AOC & 42016 & 0.012\%  \\ 
\hline
kylegriffin1 & 26270 & 0.007\%  \\
\hline
BarkyBoogz & 25503 & 0.007\%  \\
\hline
SethAbramson & 24055 & 0.007\%  \\
\hline
MeidasTouch & 23633 & 0.007\%  \\
\hline
\end{tabular}
}
\end{table}

\begin{figure}[!tbp]
  \centering 
  \subfloat[Parler]{\includegraphics[scale=0.25]{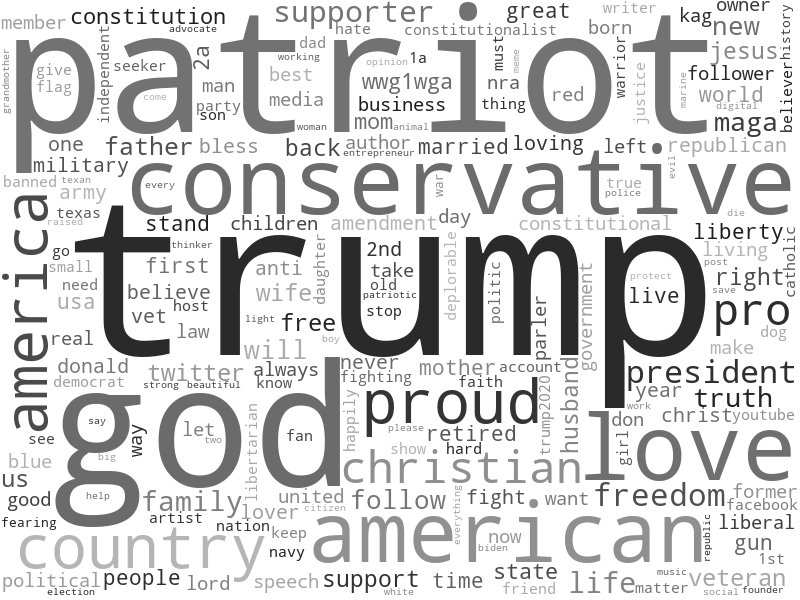}}
  \hfill
  \subfloat[Twitter]{\includegraphics[scale=0.25]{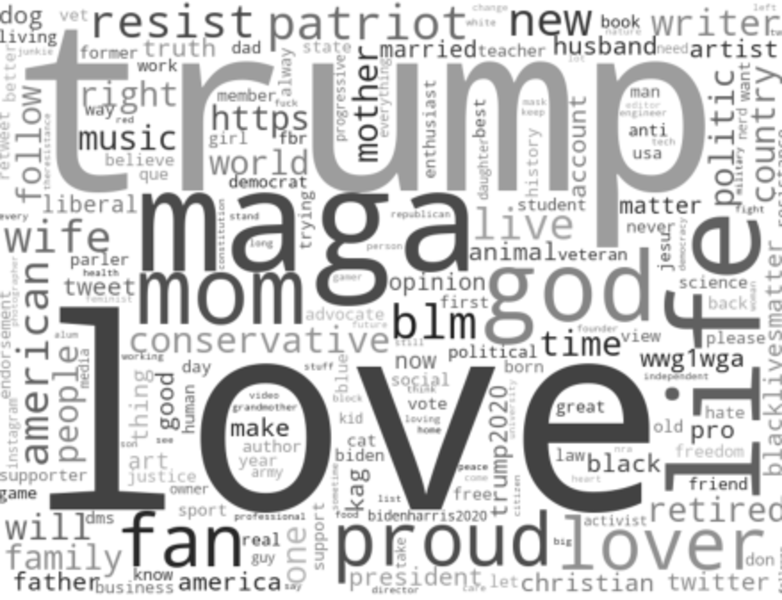}}
  \caption{Wordclouds of user bios. Twitter presents some diversity with presence of terms like \textit{love}, \textit{mom}, \textit{trump}, whereas, Parler only contains conservative right-wing terms same as those echoed by the attacking mob.}
  \label{fig:bios}
\end{figure}

\begin{figure}[!tbp]
  \centering 
  \def\svgscale{0.5}
  \subfloat[Parler]{\includegraphics[width=0.5\textwidth]{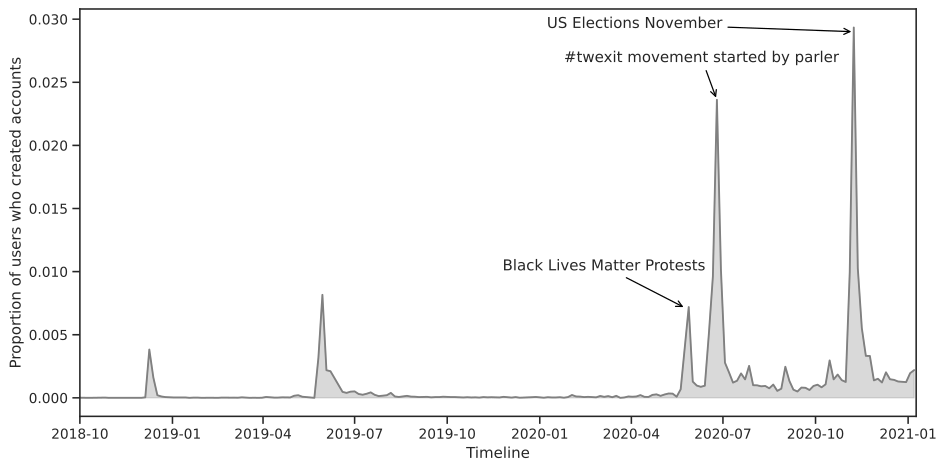}}
  \hfill
  \subfloat[Twitter]{\includegraphics[width=0.5\textwidth]{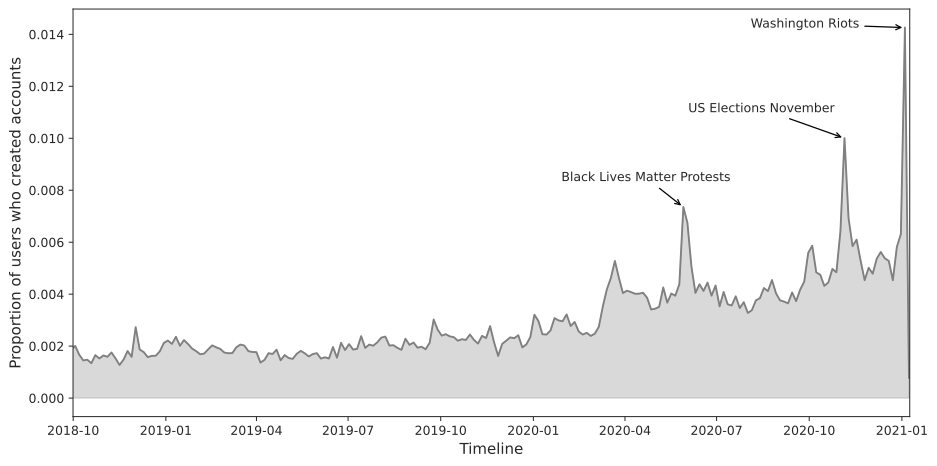}}
  \caption{Proportion of accounts created over time. All the user sign-up spikes on Parler are during a political event, explaining a highly polar userbase. In comparison, Twitter have a steady rate of sign-ups with less prominent peaks during political event. However, A large rate of sign-up is observed on Twitter during the time of riots.}
  \label{fig:account_age}
\end{figure}

Finally, to garner a sense of user-profiles we generate a word cloud of user bio and time of joining the platform shown in Figure~\ref{fig:bios} and \ref{fig:account_age} respectively. Though results shown by the word cloud, do not show a clear separation like our other plots, mainly because of a heavy presence of \textit{trump} and \textit{maga} in Twitter bios which are generally associated with conservatives. However, users on Twitter indicate some signs of diversity with presence of terms like \textit{love}, \textit{life}, \textit{mom} and \textit{fan} versus that of Parler which is only populated with terminology associated with conservatives. 

% \begin{figure}[!tbp]
%   \centering 
%   \def\svgscale{0.5}
%   \subfloat[Parler Content Trends ]{\includesvg[width=0.5]{popular-body-terms-by-post-parler.png}\label{fig:p_term}}
%   \hfill
%   \subfloat[Twitter Content Trends ]{\includesvg[width=0.5\textwidth]{twitter_top_content_normalised.png}\label{fig:t_term}}
%   \caption{Top 20 Terms Used}
%   \label{fig:term}
% \end{figure}

While analysing the user joining date, we observed three distinct peaks in Parler. From left to right in the graph, these peaks correspond to the black lives matter protests from May-June 2020, the \#twexit movement started by Parler during July 2020 and the US presidential elections during November 2020. In twitter data, we observe comparatively smaller but prominent peaks during the Black lives matter and November election. However, the outlier is the steep peak 6th of January 2021 during the Capitol hill riots. To understand the effect and role of these users, further analysis is required. Lastly, it is also interesting to observe that outside the three peaks caused by political events, user sign-ups on Parler are minuscule, explaining the highly political nature of users on Parler.

\subsection{How organic was the traffic generated?}
\label{ctm}
We calculated the Coefficient of Traffic Manipulation (CTM) \cite{nimmo2019measuring} for frequently occurring hashtags on both platforms. CTM is a relative metric to measure how much traffic of a given hashtag has been manipulated. Equation~\ref{eqn:ctm} provides an mathematical representation of CTM.

\begin{equation}
\label{eqn:ctm}
    C = \frac{R}{10}+F+U
\end{equation}
Here, for a given hashtag $t$:-
\begin{itemize}
    \item $C$ is Coefficient of Traffic Manipulation for $t$.
    \item $R$ is percentage of $t$ traffic created by reposts.
    \item $F$ is percentage of $t$ traffic created by top fifty users.
    \item $U$ is average number of posts per user for $t$.
\end{itemize}

\begin{figure}[!tbp]
  \centering 
  \def\svgscale{0.5}
  \subfloat[Parler]{\includegraphics[width=0.485\textwidth]{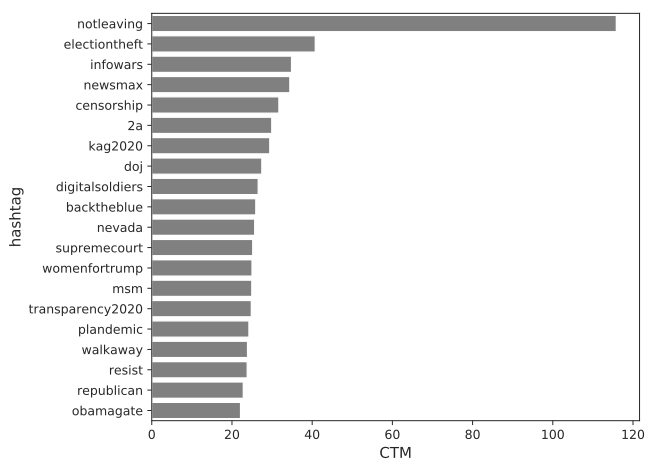}\label{fig:5a}}
  \hfill
  \subfloat[Twitter]{\includegraphics[width=0.515\textwidth]{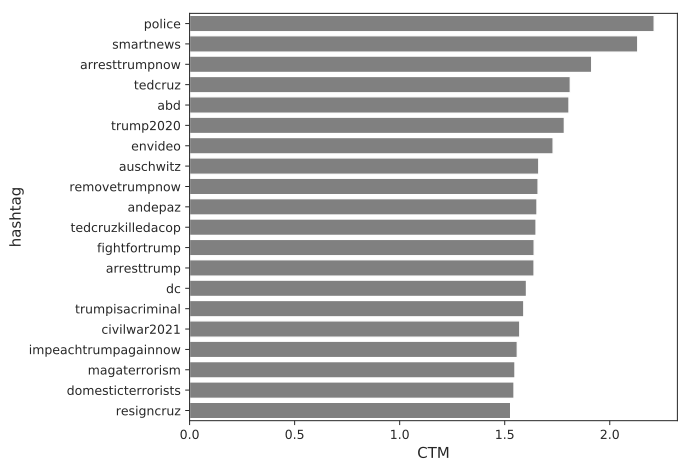}\label{fig:5b}}
  \caption{CTM on popular hashtags. Traffic on Parler is disproportionately more manipulated as compared to that on Twitter where all tags are within the range of organic traffic. Highest CTM score on Parler is $46$ times higher than that of Twitter.}
  \label{fig:ctm}
\end{figure}

Using equations \ref{eqn:ctm}, we find the value of $C$ for most popular hashtags on Twitter and Parley shown in Figure~\ref{fig:ctm}. CTM for the most manipulated hashtag on Parler is approximately $46$ times more than the most manipulated hashtag on Twitter, indicating a large extent of traffic manipulation on Parler. A similar trend can be seen for all other frequently occurring hashtags too. Though CTM is a relative scale \citet{nimmo2019measuring} in there report anecdotally found that CTM for organic traffic generally was less than 12. Judging on that scale along with the abundance of pro-riot content observed in our previous experiment indicates the presence of a manipulative rigged election agenda on Parler, which eventually could have fueled into the tragedy of 6 January 2021.

\section{Conclusion}
In this report, we analyse trending traffic from Twitter and Parler, in light of the riots at the US Capitol in Washington DC on January 6, 2020. We look at the kind of content being generated and the users involved and find evidence supporting the claims that a significant proportion of traffic on Parler was in support of undermining the veracity of the 2020 US Presidential Elections. We also find the traffic on Parler to often be violent, consisting of hate speech (as a result of Parler's relaxed community guidelines) and manipulated as indicated by CTM. On the other hand, Twitter users majorly used the platform to condemn the incident and its perpetrators and show their disdain towards President Donald Trump for his actions. An in-depth analysis of these users' networks and their activity across the two websites are required to understand better the role of social media in incitement and planning of these riots. We are continuing to analyze this data for more insights. 

\begin{acks}
Hitkul funded by TCS Research Fellowship.
\end{acks}

\bibliographystyle{ACM-Reference-Format}
\bibliography{sample-base}

% \appendixf

\end{document}